# Threat Perception Modulation by Capturing Emotion, Motor and Empathetic System Responses: A Systematic Review

E. M. Jacobs, F. Deligianni., and F. Pollick

**Abstract**— The fight or flight phenomena is of evolutionary origin and responsible for the type of defensive behaviours enacted, when in the face of threat. This review attempts to draw the link between fear and aggression as behavioural motivations for fight or flight defensive behaviours. Hence, this review intends to examine whether fight or flight behavioural responses are the result of fear and aggression. Furthermore, this review investigates whether human biological motion captures the affective states associated with the fight or flight phenomenon. This review also aims to investigate how threat informed emotion and motor systems have the potential to result in empathetic appraisal modulation. This is of interest to this systematic review, as empathetic modulation is crucial to prosocial drive, which has the potential to increase the inclination of alleviating the perceived threat of another. Hence, this review investigates the role of affective computing in capturing the potential outcome of empathy from threat perception. To gain a comprehensive understanding of the affective states and biological motion evoked from threat scenarios, affective computing methods used to capture these behavioural responses are discussed.

A systematic review using Google Scholar and Web of Science was conducted as of 2023, and findings were supplemented by bibliographies of key articles. A total of 22 studies were analysed from initial web searches to explore the topics of empathy, threat perception, fight or flight, fear, aggression, and human motion. Relationships between affective states (fear, aggression) and corresponding motor defensive behaviours (fight or flight) were examined within threat scenarios, and whether existing affective computing methods are succinct in capturing these responses, identifying the varying consensus in the literature, challenges, and limitations of existing research.

——————————— ◆ ———————————

## 1 INTRODUCTION

THE goal of this review is to identify the relationship between fear and aggression affective states and the fight or flight defensive behaviours. Our second aim is to investigate whether these human motion defence responses accurately portray and capture these affective states. Thirdly, this review intends to investigate what threat perspective factors promote empathetic responses in observers. With emphasis on affective computing methods, this systematic review intends to explore existing methods used to evoke these aforementioned affective states and to measure corresponding motor states and prosocial actions. Our ability to comprehensively understand the relationship between emotion and motor systems, as modulated by threat perception, is dependent on the affective computing methods used to evoke and capture these states. Furthermore, empathetic appraisal is of great relevancy to threat research, as empathy plays a crucial role within prosocial behaviours. Hence, in order to facilitate prosocial behaviours towards the vulnerable, a greater understanding of the relationship between threat perception and empathy ought to be achieved.

Unfortunately, this systematic review revealed that literature investigating these components is scarce, with focus on behavioural responses in animals, phobics and outgroup threat. Furthermore, it is only in recent years, that researchers have begun to utilise the advancements of affective computing to measure and evaluate the extensive relationship between emotions and motor systems [1]. The link between motor empathy and fear defensive responses, is often substantiated by studies investigating the absence of regular emotion appraisal, as observed by the lack of motor empathy and fear responsivity being found in youth with atypical emotion recognition [2]. Consequently, this review aims to summarise the literature that has measured the presence of defensive behaviours, affective states and empathetic appraisal, within interpersonal threat. Thus, intends to facilitate the understanding of interpersonal threat perception, its impact on fear, aggressive, empathetic affective states, and motor responses. As a result, existing, and more novel methods of measurement

————————————————

- *E. M. Jacobs is with the SOCIAL AI CDT, a collaborative center for doctoral training between the University of Glasgow and UKRI, G12 8QQ. E-mail: 2694507j@student.gla.ac.uk.*

- *F.Deligianni is with the College of Computing Science, University of Glasgow, G12 8QQ. E-mail: fani.deligianni@glasgow.ac.uk*

- *F. Pollick. is with the Centre for Social Cognitive and Affective Neuroscience, University of Glasgow, G12 8QQ. E-mail: frank.pollick@glasgow.ac.uk.*

will be reported within this review, to provide a comprehensive scope on the threat perception literature. This review is relevant to behavioural and human-computer interaction scientists, who intend to investigate threat perception's impact on eliciting affective states and motor responses, while expanding their insight on the best methods of measurement to use.

Our systematic review portrays a novel contribution to the maturing body of threat perception and affective computing research, through the provision of a comprehensive synthesis of the existing literature on the relationship between threat perception, affective and motor systems. This review was conducted rigorously, through of PRISMA methodology to examine, select and analyse relevant studies [3]. Hence, this systematic review offers unique insights into this complex field of emotion and threat research, providing valuable insights for future practice within this field.

## 2 CONCEPTS
### 2.1 Fear

Fear is defined as a biologically basic emotion, to promote one's own survivability. However, the role of fear is conceptualised differently by varied emotion models. Different emotion theories of fear have attempted to characterise the different variations of fear. For instance, evolutionary fear is characterised as an instance of a basic survival system [5]. Meanwhile, modular models of emotion view fear as either phobic, the reflection of the operations within brain modules, or a response to pain, predators, and conspecific aggressors in one's environment [5]. The State Affect-Related theory depicts emotions discreetly which provokes an individual to mentally focus on a specific set of actions [6]. This theory is substantiated by the experience of fear activating the fight-or-flight response [7]. Meanwhile, the Polyvagal Theory claims that affective states influence neurobiological responses of an individual, establishing a relationship between measurable physiological phenomena like heart rate and the affective state of an individual [8]. Hence, fear is often quantified by physiological arousal and reflected in fight or flight behaviours.

Fear is a central state of an organism, shaping one's conscious experience and fear behaviours that emerge circumstantially [9]. As a result, fear is a prevalent affective state in our ever-changing, unpredictable society. Fear promotes adaptive behaviours to increase one's survivability, priming individuals to detect varied threats within their surroundings. The innate nature of fear was honed by evolutionary pressures, hence, prior experience with the feared stimulus is not required to evoke defensive responses [10]. Specialised neural circuits have evolved over time, enabling individuals to detect danger within split-seconds [10]. Context-dependency of fear is observed in terms of eliciting circumstances (fight or flight), type of threat present (predator or conspecific) and predatory imminence, allowing the individual to enact voluntary and involuntary behaviours such as active defence, risk assessment movement inhibition and physiological responses [9]. Passive threat and change in one's environment can also promote feelings of fear. Defensive behaviour intensity is commonly modulated by threat distance, as proximal threats require more urgent responses, unlike distal threats [10]. As a result, fear is context-dependent on factors like distance and stimulus types, linking stimuli to flexible patterns of behaviours and can exist prior to or after the eliciting stimuli [9]. Consequently, fear enables individuals to quickly act upon threat by prioritising its urgency and increasing the execution of survival behaviours.

### 2.2 Aggression

Beyond fear, as a defence strategy, an individual has the propensity to act aggressively in the face of a threat. Aggression is the result of the combination of neural, endocrine, and behavioural mechanisms, which have phylogenetically evolved over time to accommodate split-second decisions and behavioural responses [11]. Aggression is prevalent as a response to impending threats and can be categorised into either reactive aggression or instrumental aggression [12]. Reactive aggression is defined as a response drawn out of provocation, while instrumental aggression is caused by feelings of anger [12]. Hence, the likelihood of an individual responding aggressively to threat, is dependent on both cognitive inhibition of aggressive acts and the predetermined tendency of one to react defensively [13].

Instrumental and reactive aggression are both mediated by the presence of integral emotions, which are emotions that evoke strong motivations and behaviour with little thought [14]. For instance, individuals may behave aggressively instead of passively, when provoked via physical altercation [15]. Hence, defence mechanisms are behavioural outcomes shaped by integral emotions. This phenomenon is explained by hostile attribution bias, whereby the attribution of hostile intent towards the attacker increases with the increase of hostile emotions experienced [16]. Consequently, hostile intent promotes aggressive behaviour in actors, despite potential consequences [17]. Hence, the role of emotion recognition in analysing hostile intent, is crucial in the decision to engage in a fight. Hence, integral emotions may bias defensive responses, resulting in aggressive behaviours towards provocations.

Despite anger being associated with instrumental aggression, threat events are more than capable of evoking both anger and fear [18]. Furthermore, protective forms of aggression are often the result of fear-motivation [19], which displays differently from anger-induced aggression, in emotional expression, biological motion and neural bases [20]. Hence, the next section of this review, attempts to illustrate the relationship between fear and aggression, within fight or flight behaviours.

## 2.3 Fight or Flight

Defensive behaviours elicit motor responses, as observed by fleeing (increasing distance from danger) and fighting (dissuading provocative action) [21]. The fight or flight phenomena is theorised to have been borne out of survival-related abilities, like fear and aggression circuits, whereby ecological rules govern fixed defensive responses [13].

There is limited human literature investigating the relationship of fight or flight behaviours within active threat. This handicaps the existing literature, as when threats are presented retrospectively, individuals might not be forced into the split-second dichotomy of fight or flight behaviours. Furthermore, studies that do investigate this phenomenon, tend to review this complex relationship on a surface level, whereby aversive behaviours associated with flight behaviours are assumed to be devoid of aggressive behaviours or emotional states. For instance, studies investigating motivational approach tendencies within a threat scenario, have attributed strong approach motivation and startle response inhibition to high trait anger [22] These studies often analyse threats from a retrospective perspective, asking for reflections on past emotions of fear and anger, rather than the present emotions elicited from an active threat. Such findings are often not substantiated with psychophysiological evidence like brain neuroimaging or affective measure; thus, negative affects are investigated solely by one's willingness to approach or avoid.

Furthermore, the relationship between fear and anger within threat scenarios, are often quantified by the fight or flight phenomenon. By measuring fear through startle responses, fearful behaviour is quantified by aversive behavioural responses, over aggressive confrontational behaviours [22]. Meanwhile, approach motivation is associated with aggression and results in the inhibition of startle responses. However, fear or aggression as depicted by fight or flight phenomenon is not mutually exclusive, as observed by the emergence of defensive aggression [23].

Defensive aggression is rooted in fear with the sole purpose of eliminating threats within one's proximity by reducing arousal and maintaining safety [24]. Limited clinical studies have shown that the effect of fear can result in the precipitation of aggression in response to threat [25]. In contrast, threat literature often views fear and aggression responses as not mutually exclusive, whereby if an individual interprets a situation as threatening, it is believed an individual would opt for threat avoidance behaviours over approaching the threat [13]. Despite reactive fear being frequently linked with flight behaviours, over fight behaviours, individuals can engage in pre-emptive defensive aggression when encountering threats [26].

Regardless of the imposed behavioural binary from fight or flight, the conditional effect of emotions colours the associations between behavioural intentions and actions [27]. This effect is observed within behaviours evoked from defensive aggression. This results in individuals potentially undervaluing, overvaluing, and even neglecting the risks and costs of the desired action [26]. Consequently, the best way to determine whether an individual resorts to aggressive-related or avoidance-related behaviours, is through observing neurobehavioural responses to provocations, as quantified by affective and motor states [13]. Hence for the sake of this review, defensive behaviours of fight or flight, will be examined in relation to feelings of fear and aggression, evoked from the presentation of threat.

## 2.4 Threat Perception

Darwin's evolutionary theory states that adaptive motor behaviours are primed to react to emotion-eliciting contexts due to closely interacting emotion and motor systems [28]. Furthermore, the presence of threatening stimuli can be signalled interpersonally via social and negative emotional cues to mediate motor responses. Negative emotional cues include aggression or fear, which can spur the observer into fight or flight [29]. Negative emotional cues are thought to be processed by the posterior cerebellum, resulting in threat preparation mechanisms being based on prediction mechanisms, which are informed by emotional cues [29]. For instance, an individual expressing a fearful expression may be signalling imminent danger while an aggressive expression may indicate themselves as the present threat. Upon witnessing both expressions, the observer is informed of two different situations that can regulate their defensive behaviours. Hence, when these cues are expressed by other agents in a social situation, these emotional signals can convey imminent danger within one's surroundings to the perceiver, alerting the observer to take the appropriate defensive action [29]. In some cases, the observation of another's response to threat, can evoke empathetic behaviours in observers. This systematic review intends to substantiate the complex relationship between fear and aggression through the fight or flight phenomenon the face of threat. Furthermore, this review intends to explore how empathetic appraisal can be facilitated within an observer to another experiencing threat.

## 2.5   Empathy

Empathy is defined as experiencing other-oriented emotional responses, which are evoked and congruent with the perceived emotion of another in need [30]. Other sources of empathy include perspective-taking and concern for another's welfare [31]. Similarity of experience is believed to contribute towards empathy, whereby a sense of unity can be fostered upon observation of similar responses and mutual life events, regardless of cultural differences [32].

Empathy is the mutual sharing of experiences and is facilitated by emotional congruence with another individual in need [30]. Empathy operates at different levels of generality, whereby similarity of experiences can range immensely from a precise match to a common denominator [32]. As a result, to cognitively empathise with another individual, one need not undergo the exact same experience as the other. But rather, empathy can be evoked from generalised experiences that resemble a situation that the empathiser has experienced prior [32].

Empathy is believed to promote emotion regulation during conflict, by providing individuals with an empathetic lens to reframe the current situation, thus, understanding another's perspective [34]. Emotional regulation provides the opportunity to modulate how an emotion is experienced and the consequent action that emerges from the emotion [34]. Outcomes of empathetic emotion modulation include prosocial behaviour and social bonding, by compelling individuals to aid others [35]. The interaction of multiple neural circuits related to emotional, motivational, cognitive, and behavioural functions are responsible for these prosocial outcomes [36].

However, despite the relevance of empathy towards threat perception research, research on empathy tends to investigate the deficit of empathy rather than the presence of it, with priority on the inability to respond to a victim's distress [37]. Furthermore, there is lacking research investigating the regulation of threat arousal [38], nor is there research on empathy examined from a bystander's position. This review aims to provide additional understanding on the influence of empathy on emotion regulation and threat perception.

## 3 MATERIALS AND METHODS

The systematic review presented in this article was conducted in three steps – Planning, Selection and Critical Analysis of Results, as suggested in [39]. For Planning, a protocol was defined, and the research questions are presented in Table 1.

TABLE 1
RESEARCH QUESTIONS

| # | Research Questions |
|---|---|
| RQ1 | Are fear and aggression linked within threat? Do fight or flight behaviours result from this relationship? |
| RQ2 | Can biological human motion portray and capture measurable affective states from the fight or flight phenomenon? |
| RQ3 | What factors of threat perception modulates empathetic responses and how are these measured by affective computing? |
| RQ4 | How do empathetic appraisals result in prosocial behaviour, and has this been measured by affective computing? |

A computerised search was undergone on Google Scholar and Web of Science for relevant articles. A combination of keywords was used in the searches. The first combination of key terms was: ("biological motion", "empathy", "aggression", "fear", "virtual reality"). The second combination of key terms included ("threat", "empathy", "fight or flight"). The third search included, ("threat", "vicarious fear"). The fourth search included the words ("defensive aggression", "threat", "fear"). The next search included ("threat", "fear", "empathy"). The fifth search was a combination of the key terms ("human motion", "perspective-taking", "empathy", "virtual reality"). The results that were gleaned from this conducted search was as of January 2023.

In the Selection step, the following inclusion (I) and exclusion (E) criteria was adopted:

(I1) Articles that address the impact threat perception has on fear, aggression, and empathetic appraisals.

(I2) Articles published and fully available on scientific databases.

(E1) Studies not published in English.

(E2) Studies investigating non-human species.

(E3) Studies that investigated and utilised non-neurotypical subjects.

(E4) Studies that did not undergo peer-review.

A summary of this criteria is depicted by Fig. 1.

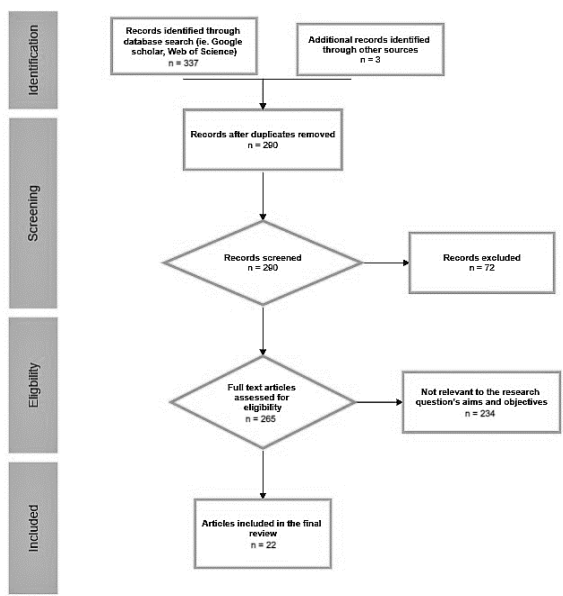

**Fig. 1.** Overview of the systematic review process

Additionally, the following quality criterion was defined: works that comprehensively investigate and measure the effect of threat perception on the relationship between emotion and motor systems. Furthermore, studies that examined vicarious threat learning through embodied perspective taking and virtual reality (VR) methods, were included. This criterion was applied in the Extraction step, during the full reading of the text. Studies that met both the inclusion criteria and the quality criterion were included in Table 2.

TABLE 2
OVERVIEW OF EMPIRICAL STUDIES REFERENCED

| # | Citations | Physiological Measurements | Results |
|---|---|---|---|
| 1 | [13] | Startle response, EMG measurement, Reaction time, MRI recording | High emotional reactivity to threat suppresses recruitment of the mentalizing network |
| 2 | [25] | State fear, blood pressure, and heart rate | Attention bias to threat was positively associated with in vivo aggression. Greater attentional bias to threat was associated with less escape behaviour, within fear conditions |
| 3 | [28] | Heart rate variability, posturographic measurements | Posturographic analyses showed that angry faces resulted in significant body sway reductions. This reduced body sway was accompanied by bradycardia and correlated significantly with subjective anxiety. |
| 4 | [38] | SCR, heart rate | Individuals with higher trait resilience and individuals with higher resting heart rate variability showed more regulation in terms of their subjective arousal experience. |
| 5 | [40] | Heart rate | Defensive states that drive active escape from danger may facilitate decisions to help others, by engaging neurocognitive systems |
| 6 | [41] | NA | There is a disconnection between aggressive behaviour and negative affective states accompanying avoidance motivation within the behavioural inhibition system. |
| 7 | [49] | Transcranial magnetic stimulation and EMG | Extremely rapid bilateral modulation of the motor cortices occurs upon seeing emotional bodies, with stronger suppression of motor readiness when seeing fearful bodies. |
| 8 | [61] | Skeleton and joint data | Method achieves higher recognition accuracy on emotion expression classifications. |
| 9 | [64] | Kinect, head tracking, ECG, galvanic skin response and EMG | High inter-rater agreement of aggression and dominance variance was observed for actors, who engaged in overt behaviour, over the naive participants. |
| 10 | [74] | NA | Greater empathic responses towards a virtual human's pain when they present natural postural oscillations |
| 11 | [77] | 3D motion features and temporal features | The emotion recognition system achieved recognition accuracy of 90.0% during walking, 96.0% during sitting, and 86.66% in an action-independent scenario. |
| 12 | [78] | Reaction time | Findings demonstrated that approaching threats elicit freezing in humans. |

| | | | | |
|---|---|---|---|---|
| 13 | [84] | EEG | More expected pain of others induced stronger empathic responses as observed by EEG signals. | |
| 14 | [89] | fMRI | Amygdala activity was mediated by emotional intensity and has a role beyond disambiguating stimuli. | |
| 15 | [92] | EEG | Results indicate that specific social contexts can modulate neural responses to observing another's pain. | |
| 16 | [93] | Virtual Reality Paradigm | After embodying a female victim, offenders improved in ability to recognise fearful female faces and reduced their bias towards recognising these expressions as happy. | |
| 17 | [95] | fMRI and Virtual Reality Paradigm | First-person perspective of a virtual violent situation impacts emotion recognition through DMN activity modifications. | |
| 18 | [108] | Baseline physiological signals (HRD) from Virtual Reality Paradigm | Embodiment from a child's perspective during conflict in VR impacts emotion recognition, physiological reactions, and attitudes towards violence. | |
| 19 | [115] | fMRI | Neural correlates of brain regions and self-reported feelings of compassion. | |
| 20 | [117] | SCR | Observation of another's threat reactions can recover associations previously shaped by direct, firsthand aversive experiences. | |
| 21 | [118] | SCR | Results confirm that vicarious threat learning can be evaluated experimentally, but empathy does not amplify this process. | |
| 22 | [122] | Head tracking data, physical behaviour recordings, proxemics | Compassion and tendency to experience personal distress predicted empathetic concern and proxemic behavior (gaze orientation and degree of interpersonal distance) to virtual peoples in distress | |

## 4 THREAT PERCEPTION, AFFECTIVE STATES AND DEFENSIVE BEHAVIOURAL RESPONSES

### 4.1 Fear and Aggression Influencing Threat Perception to Modulate Fight or Flight Behaviours

Fight or flight defence mechanisms are dependent on perceived threat imminence and are coordinated by organised neural networks that have evolved over time, resulting in split-second decisions that aim to avoid or deter impending threat [40]. As a result, negative stimuli can rapidly trigger fight or flight motor responses [29]. Emotion cues such as fear and aggression result in motor defensive behaviours due to the extensive relationship between emotion perception and action systems. The neural function for fear perception is characterised by rapid and automatic processing, while anger perception is dependent on conscious processing [41]. Due to the rapid nature of automatic processing, fear is the most strongly associated emotional state with defensive behaviours, controlling when an individual resorts to fleeing or freezing [42]. Threat perception motivates the adaptability of survival behaviours in humans, influencing fight or flight behaviours [25].

The relationship between neural structures and threat perception is so prominent, that defensive motor reactions, are influenced through the passive observation of emotionally aversive and arousing stimuli [43]. The interconnected neural architecture involved in threat detection consists of cortical and hippocampal circuits, towards attention systems like the amygdala, striatum, and defensive circuits within the midbrain [44]. Hence, due to survival pressures, neural networks are dedicated to processing fear and threat, in attempts of balancing homeostatic environmental threats and outsmarting predators [44]. These neural networks are theorised to operate as a threat orienting network, by assessing threatening stimuli before guiding motor actions crucial to minimise potential harm [44]. Despite the speculated prevalence of cognitive structures in processing fear and aggression within an observer, there is limited neuroimaging literature exploring threat perception paradigms. This is due to most findings within the literature being rodent focused despite the cross applicability of these findings to human research [44]. Thus, human research within interpersonal threat paradigms is limited.

Various brain regions have been speculated to play a central role towards aggression, fear, and threat perception responses. Subcortical regions of the brain promote reflexive and subconscious threat responses, like physiological reactions and defensive actions [25]. Meanwhile, cortical regions mediate cognitive processing of threat information, via prior learning and ongoing physiological sensations, promoting feelings of conscious fear [45]. Regarding aggression, the stimulation of cortical regions such as the periaqueductal grey (PAG), amygdala and hypothalamus have been theorised to comprise a rage circuit [46]. Similarly, when within a threat's proximity, activity in the PAG is observed [47], with its stimulation resulting in fear-induced symptoms of terror and the desire to escape the situation [48].

Fear emotive cues are perceived differently from other emotive cues, as observed by distinguished neural responses to differing emotive stimuli. A study investigated cortical mechanisms involved in the implementation of motor and emotional responses, in the presence of fear-evoking cues [49]. The hypothesis that the observer's motor cortex readily suppresses motor readiness towards neutral stimuli, when observing fearful body cues, was measured by corticospinal excitability. Pictures depicting emotive body language were observed, with fearful images resulting in intracortical suppression, unlike happy images [49]. Findings indicate that when perceiving emotional bodies, rapid bilateral modulation of motor cortices occurs, while fearful bodies evoke stronger motor readiness suppression [49]. Hence, fear driven emotive bodies are not only cognitively distinctive from happy emotive bodies from an observer's standpoint, but they also promote motor readiness for defensive actions. As a result, these findings substantiated the fact that defensive motor systems are intrinsically linked to the perception of emotion cues. Thus, providing neurophysiological support for the evolutionary relationship between emotional perception and to action systems, enabling the quick mobilisation motor reactions [49].

Fearful body cues are prioritised in perception, as observed by distinctive brain regions sensitive to fear cues and their neural processing differences. Brain regions sensitive to fear have been identified and support the hypothesis that fearful body cues are perceived differently. For instance, both the extrastriate body area (EBA) and the fusiform body area (FBA) have been identified as central to expression recognition [50]. However, fear modulation has only been found in the EBA, unlike the FBA, despite this difference not being found in other emotions [50] This difference in fear sensitivity is believed to be due to the EBA playing a central role between perceptual and motor processes [51]. Furthermore, magnetoencephalographic measurements have identified fearful bodies being processed 80ms earlier than neutral bodies in the right parietal cortex, providing evidence for the emotion-attention-action link within the dorsal visual stream, which is crucial for motion analysis [52. The preference for processing fearful stimuli was observed by the faster processing of fearful body expressions than neutral or happy expressions, by event-related potentials [53].

### 4.2 Fear and Aggression as Measurable Affective States, Derived from Behavioural Responses

Affective states are psycho-physical phenomena measuring emotional states through valence and arousal. Valence is the evaluation of stimuli on a positive-to-negative scale, while arousal is dependent on the excitement of the sympathetic nervous system [54] According to the elementary arousal model, an increase in physiological arousal, heightens the intensity of emotional responses [55]. Consequently, behavioural responses are best measured from physiological responses when affective states are induced. By quantifying behavioural responses from affective states, affective computing can provide further insight into threat perception.

By this definition, fear is an affective state, as it comprises of elevated physiological arousal, from the subjective experience of fear [56]. Furthermore, exposure to feared stimuli induces physiological arousal and subjective fear responses [57].

Meanwhile, heightened physiological arousal increases the likelihood of aggressive fight behaviours, within anger appraisals [58]. Aggression has been positively correlated with the expression of anger [59], due to anger being an emotion and aggression being a behavioural response [560]. Despite aggression not being an emotion, it is an affective state derived from anger, thus, is a quantifiable behavioural action that is observable and measurable. Furthermore, it is to be noted that aggression and anger are highly conflated within the literature, due to the high inter-relatability of these two constructs [60].

### 4.3 Limitations of Existing Measures of Behavioural Responses

As emotion research continues to advance, traditional emotion and motor recognition methods become increasingly restrictive. Traditional physiological methods of emotion recognition require subjects to wear a series of relevant sensors for data collection, imposing restrictions on naturalistic behaviours and acquisitions, compromising the accuracy of emotion recognition for real world settings [61]. Traditional physiological measurements include indirect measures of the autonomic nervous system, like fear-potentiated startle as measured by electrocardiogram (ECG), skin conductance responses (SCR), eye tracking and measurements of direct brain activity from functional magnetic resonance imaging (fMRI) [62].

The measurement of human behaviour brings about its own methodological complexities, due to the multidimensional nature of social interactions. Threat interactions are difficult to control due to their nonlinear and multidimensional nature [63]. Furthermore, inducing fear and aggression in an experimental context, brings about ethical concern, due to the lack of experimenter control over potential breaches of the participant's safety. One traditional approach is the use of human observers to observe, code and record evaluations of behaviours and emotion states from social interactions. However, coding schemes utilised by the observer are limited by observable physical aspects of the interaction, such as mutual gaze occurrences, or generalised impressions of the overall interaction [63]. Additionally, the mismatch of subjective responses being captured by rigid coding questionnaires reduces the richness of the behavioural data captured. The quantitative nature of these coding schemes also hinders the reproducibility of the experiment, due to the lacking ability to control and report subjective variables consistently. Where evaluative approaches are qualitatively rich, they come at a cost of time and resources required to maintain the richness of the data, especially over longer durations of time [63]. Furthermore, the ecological validity of human coding is further limited by the subjectivity of the observer's interpretations [63].

Meanwhile, the study of defensive behavioural motion from real-life provocations are limited, due to the lack of conflict predictability occurring in naturalistic scenarios. Hence, it is difficult to firstly, utilise real-world paradigms to control exposure to interpersonal provocation for experimental study and secondly, to accurately measure human behaviour, comprising of defensive motor responses and affective states. in social interactions.

### 4.4 Benefits of VR as a Methodology to Portray and Capture Measurable Affective States

To combat the issue of ecological validity and replication issues, VR paradigms were developed, to offer a three-dimensional virtual world for measuring behaviours within social situations [64]. VR presents the opportunity to examine emotional perception and behavioural tendencies through the manipulation of experimental conditions. It provides the field of affective neuroscience with an effective tool to study affective loops, under controlled conditions in settings where real-life manipulation would be impossible to control or unethical [65]. For example, the study of prosocial behaviours when confronted by a violent incident can be studied, without bringing real harm to participants [66]. The potential of VR creating naturalistic paradigms can create ecologically valid stimuli that are representative of real-world situations. This method allows experimenters to seamlessly observe and record behavioural tendencies that surface within these virtual environments, without breaking the immersion of participants. Additionally, VR experiments provide a greater degree of experimental control, which is essential when exploring highly emotive situations, such as interpersonal threat.

Traditional methods of physiological measurement are commonly used within the existing literature to supplement VR observations of a participant's affective state. For instance, within a VR simulation, changes in postural and autonomic behaviours towards virtual avatars displaying different emotion was measured by postural displacement, and by skin conductance [67]. Similarly, heart rate variance was measured as a basis for continuous arousal a key feature of fight or flight body mobilisation [68, while participants were to physically navigate threats within an immersive virtual environment [38]. Where the sole use of physiological measurements falls short, the combination of VR derived measurements and physiological measurements paints a wholistic view of captured biological motion and their corresponding affective states.

VR complements physiological measurements by improving the validity of ethological physiological measurements. This is observed by proxemic measurements, a common example of in vivo measurement from social interactions without having to hook individuals up to motion capture monitoring equipment or constrict them to the confining space of a scanner. As a result, VR can use motion analysis to produce valid measurements of behavioural responses, without restricting the movement of participants.

### 4.5 VR Utilising Motion Analysis to Analyse Displays of Empathetic Behaviour

Proxemic behaviour was found to be influenced by displays of empathy, as observed by the mediation of one's physical distance to individuals in need. Research has indicated that prosocial behaviours are positively associated with less physical distance between individuals [69]. Prior to VR, motion, and eye tracking were commonly used to record one's gaze, orientation, and position continuously and precisely over the course of an interaction. VR provides the opportunity to measure subjective elements like continuous or covert mutual gaze and interpersonal distance. Within VR, proxemic behaviours such as the movement path of an individual and the head orientation while moving within the VR environment are measured.

VR measurements provide more insight than the excessive amount of tracking data from traditional motion-tracking. The latter is not only challenging to reduce into a manageable form, but the common reduction to mean or minimum interpersonal distance may wash out meaningful results, making the data sensitive to noise [63]. Thusly, VR proxemic measurements attempts to combat the reduction in interdependent measures of interpersonal distance from one's gaze orientation, unlike traditional motion-tracking. The combination of VR and motion tracking presents the ability to study social interaction, while systematically analysing how kinematics differ between emotional scenarios [71].

Furthermore, VR remedies the difficulties of measuring prosocial behaviours in experimental contexts, due to the varying situational inclinations promoting consistency in confederates' behaviour across trials. Conditions such as the environment that the virtual environment embodies, and the behaviour of a virtual confederate can be controlled and kept consistent between trials. For example, individuals can feel more inclined to act prosocially when confederates display increased eye-contact [71] or when situational cues signalling to respect social norms are visible within the environment [72]. As a result, participants can be encouraged towards enacting prosocial behaviours when the situation they find themselves within is controlled entirely to evoke an empathetic response in observers. This method of online data acquisition presents advantages of measuring subjective social behaviour, to traditional methods, as observed by self-reported proxemic measures and passive role-playing scenarios. Hence, research suggests that VR could provide a more valid and nuanced method of measuring proxemic behaviours than traditional motion capture methods, by accounting for critical measurements beyond interpersonal distance, such as gaze behaviour [63]. Increasing research has indicated the benefits of using VR to capture emotions and threat recognition from the measurement of human motion.

**4.6 VR Capturing Emotion and Threat Recognition from the Display of Human Motion**

The modulation of emotions evoked from interactive virtual environments (IVE) is dependent on the degree of immersion and presence experienced from successful interactions with virtual avatars [64]. These IVEs have higher degree of emotion recognition accuracy from the perception of emotive motor behaviours, when a large range of somatosensory sensors are used to capture behavioural responses across different modalities.

Negative affects which can be measured within VR paradigms, through changes in facial expressions, body language and physiological changes [73]. To investigate the effect of emotive body language on observer physiological ratings, aggression-evoking role-play scenarios were utilised by trained actors and participants to train avatar models for aggression prevention virtual paradigms [64]. Hence, to obtain a rich analysis on negative affects, they captured overt behaviour through audio, video, motion capture and head tracking. Physiological changes were recorded by electrocardiogram (ECG), GSR and muscle activity of the biceps, triceps, and trapezius muscles (EMG). Results showed that aggression, fear, valence, arousal, and were more highly rated for overt body language being displayed, in comparison to spontaneous and sometimes covert behaviours [64]. Similarly, activation was found in the trapezius, deltoid and tricep muscles in fear perception, while the biceps displayed inhibition [50].

Hence, these findings indicate that the overt use of confrontational body language, promotes levels of fear and aggression within viewers, being central in the portrayal and detection of aggression. Furthermore, this novel study indicated the potential of overt body language being used to inform virtual agents within VR simulations, to evoke affective states of fear and aggression.

There is a lack of research utilising IVEs to capture and analyse the effect of virtual agents portraying non-communicative movements on the modulation of perceived emotions [74]. Non-communicative movements consist of idle motion and are commonly used to simulate behavioural realism in virtual simulations [74].

VR presented the opportunity to study the influence of natural postural oscillations on the perception of another's pain and the corresponding empathetic response. Realism in biological motion from idle motion was found to be important in the perception of virtual agents, promoting feelings of empathy within observers. Participants who witnessed static virtual agents who expressed pain, rated the pain observed as more intense [74]. Meanwhile, virtual agents that were animated with idle postural oscillation resulted in participants feeling more empathetic towards the pain expressed by the virtual agents [74]. These findings coincide with the fact that the human visual system is sensitive to biological motion, hence, the presence of biological motion may result in individuals identifying the virtual agent as a natural agent (i.e., human), instead of an artificial one (i.e., cartoon) [74]. This classification of virtual agents as natural agents promotes immersion of the participant, creating more validated results measured within interpersonal social contexts. This early identification stage influences latter stages of facial pain expression recognition, modulating realism and promoting higher empathetic responses to pain [74]. However, it is still unclear whether the presence of idle motion affects the participant's subjective judgements in general, as opposed to the empathetic feelings towards the pain of virtual agents [74]. Consequently, future research needs to further explore the categorisation of animated agents as humans, and whether human motion realism affects empathetic resonance of pain [49].

Multiple machine-learning methods have been proposed to classify emotions from whole-body human movements, within virtual environments. This method of emotion recognition complements and enriches existing methods of facial emotion recognition, due to the ambiguity of facial expressions in isolation [75]. Additionally, the recognition of facial expressions has been found to be modulated by the expression of the body [76]. Hence, bodily expressions help to further provide the emotional context of the situation. This inclusion paints a holistic view of emotion expression as in reality, faces cannot be viewed in isolation from the body. The recognition of emotions from motion also serves to inform behavioural responses that biases emotion perception [65]. These machine-learning networks have the potential to redefine behavioural modelling, by strengthening the relationship between computers and humans, through the automatic recognition of emotions from body movement [77].

The use of whole-body movements to extract emotion recognition data from VR, was proposed, in the face of physiological method limitations [61]. The stack Long Short-Term Memory network based on attention (AS-LSTM) was proposed for whole-body emotional recognition, to prioritise key joint point features and improve the network's accuracy in emotion recognition [61]. This proposed method managed to classify emotions such as happiness, sadness, fear, anger, surprise, and disgust, indicating effectiveness of this emotional recognition method being implemented in future VR studies [61]. Similarly, the use of whole-body expressions as a modality for emotion recognition from emotionally expressive walking sequences, can predict the intention and emotional state of individuals and encompassing movements relevant to the reflected emotional state [77]. Not only does whole-body motion enrich emotion recognition methods, but they also provide insight towards the extraction of emotional features from body movements, being used for the classification of emotions between humans and virtual agents [61].

### 4.7 The Effects of Threat Perception Captured by Absence of Biological Human Motion

The absence of defensive behaviours, portrayed through freezing behaviours, further illuminates the relationship between threat perception and biological human motion. However, there is limited literature investigating the freezing of motion and slowing of reaction times (RT) in humans, in the face of threat [78]. Consistent with the Threat Detection System theory, individuals detect and react rapidly to threat, with approaching threats evoking greater freezing responses, as opposed to stationary or receding threats [79]. The role of threat imminence was explored by investigating whether approaching threatening stimuli induces freezing of motion, when conducting a semantic decision task [78]. Findings indicated that freezing behaviours are associated with immobility, resulting in slower RT to the threatening stimuli. Furthermore, the higher the state anxiety, the stronger the freeze behaviour as operationalised by slower RT, was displayed to approaching threats, drawing a positive correlation between fear and freeze behaviours [78]. As a result, approaching stimuli predicted higher immobility when individuals displayed higher levels of state anxiety.

Social cues can be represented by facial expressions and can invoke body immobility through the elicitation of fear-potentiated startle responses. These social cues can result in attention bias towards the expressions and evoke avoidance tendencies. Angry expressions are often perceived as sources of social threat through the portrayal of dominance to induce fear in others. Despite the limited studies investigating whether freeze-like responses extend to social threat cues, social threat is thought to invoke body immobility in humans, as observed in animals [28]. To evaluate the relationship between expressions mediating motor responses, sociobiological responses to social threatening, neutral and affiliative cues (angry, neutral, and happy expressions, respectively) were analysed. Upon presentation of angry face stimuli, participants displayed reduced body sway and bradycardia, which significantly correlated with ratings of anxiety in contrast to other expressions [28]. Consequently, social threat cues of emotion expression can mediate spontaneous motor responses, as observed by freeze-like behaviours being elicited.

## 5 MOTOR DEFENSIVE RESPONSES OF FEAR AND AGGRESSION MEDIATED BY COGNITIVE FLEXIBILITY

### 5.1 Methods of Measuring and Detecting Cognitive Flexibility

Cognitive flexibility is intrinsic to adaptive behaviour, promoting survival responses over other bodily functions. Cognitive flexibility is believed to result in increased measurable arousal responses, when in the presence of threat. High cognitive flexibility results in the allocation of cognitive resources for environmental demands, priming and prioritising one's adaptive and defensive behaviours over other functions [38]. The prioritisation of defensive responses is observed by the prominence of physiological arousal in threat responses. Physiological arousal is often measured by skin conductance and heart rate variability, indicating the mobilisation of the body for defensive or offensive behaviours like fight or flight [38].

Within the literature, cognitive flexibility has also been measured by affective-swapping neurocognitive tasks. A popular method to measure cognitive flexibility is the Wisconsin Card Sorting Test (WCST), whereby participants undergo trial and error to match cards, without specific instruction as to what the sorting rule is [60]. Despite its common use, the WCST has been criticised for varying definitions and inconsistency in measurement variables, due to the various iterations of the task [80]. Another common task used to measure cognitive is the Stroop Test. Stimuli often consists of names of colours (i.e., "red", "blue") and in congruent trials, words were printed in the same colour of the word's meaning, while incongruent trials had a different colour display to the word's meaning. Participants' RT would be measured by long it took for them to match the relevant stimuli to the ink colour of the target stimulus [81]. RT is also a common. standardised method of measuring cognitive flexibility through the measurement of switch costs from affective switch tasks [82]. The switch cost is calculated by subtracting the average RT on repetition trials from the RT on affective-switching tasks [82]. Higher switch costs are indicated by higher RT, and reflect less efficient affective switching, thus decreased cognitive flexibility. Thus, the literature depicting methods of measurement for cognitive flexibility is varied.

All measures of cognitive flexibility were hypothesised to be correlated with one's arousal, with high cognitive flexibility being most effective in phases dependent on emotion regulation within prolonged threat presence. Immersive virtual environments exposing participants to either prolonged or intermittent threats facilitated measurements of physiological and subjective arousal to account for threat responses. Studies have found that RT, heart rate variability and cognitive flexibility predicted differences in task-switching flexibility, unlike skin conductance ratings, despite all these measures predicting arousal in the presence of threat [38]. Hence, affective processing flexibility predicts higher arousal responses, signalling the importance of cognitive flexibility processing in threat responses [38].

## 5.2 Cognitive Flexibility Mediating Adaptive Affective and Motor Responses in the Face of Threat

Cognitive flexibility is a crucial cognitive function that enables the switching of cognitive strategies, allowing for the adaptation of behavioural strategies in the face of threat [60]. Adaptive behaviour is the solution to ever-changing threatening situations, requiring fine-tuned, dynamic regulation of behaviour to match the environmental circumstances [38]. Hence, to engage in adaptive behaviour whilst within the presence of a threat, cognitive flexibility is crucial to hierarchise survival responses over other bodily processes [38]. Cognitive flexibility promotes the ability to spontaneously adapt through executive means, through the allocation of attentional and cognitive resources to new information, whilst inhibiting less relevant information [83]. The inhibition of cognitive processes that are not relevant to fight or flight behaviours is central to threat response, observed by the prioritisation of attention towards threat presence [38].

Cognitive flexibility prioritises the presence of threats within one's environment as individuals experience attentional biases towards threat, promoting the mediation of defensive responses through emotional regulation. Threat perception plays a crucial role in the motivation of adaptive survival behaviour, influencing defensive measures such as aggression and escape behaviours [25]. Attention is mediated by acts of learning which drive both fast hard-wired systems and top-down filtering mechanisms supported by predictions [44]. The combination of information derived from prior learning predictions and present subconscious physiological reactions require quick responses, resulting in fear sensations [84]. Furthermore, the revised behavioural inhibition system (BIS) comprises both systems and accounts for the mediation of behaviours characterised by either approach-approach, approach-avoidance, or avoidance-avoidance conflict [42]. The BIS complements fight or flight behaviours, by inhibiting behaviours reliant on top-down attention systems, and instead facilitating risk-assessment behaviour through increased vigilance [42].

Attentional bias can modulate negative affective states, resulting in the regulation of defensive behaviours. To examine the relationship between attention biases and emotion regulation of aggression and fear, an eye-tracking paradigm was used alongside the Point Subtraction Aggression Paradigm, a computerised paradigm that allowed participants to punish an opposing confederate to increase one's monetary winnings [25]. Results indicated there was a positive association between aggression and attention bias towards threat. Fear was not found to strengthen the relationship between aggression and threat, nor did it modify the relationship between aggression and attention bias. [25]. Meanwhile, the fear condition resulted in greater attention bias to threat and resulted in less escape behaviours. Consequently, fearful stimuli present in one's environment, may promote attentional fixation on it, affecting corresponding fight or flight behaviours.

These findings contribute to the sparse literature on the role of cognitive flexibility adapting affective and motor responses. However, these results may have been affected by the low ecological validity of paradigms used, due to the ethical issue of inducing threat. Emotions associated with the participant's desire to "protect" one's own funds and to "punish" a fellow confederate might not be influenced solely by fear or aggression, due to the use of monetary reward. Consequently, to investigate the relationship cognitive flexibility and affective responses such as fear and aggression, threat paradigms need to be situationally relevant to trigger defensive survival processes.

## 6 THREAT PERCEPTION INDUCING EMPATHETIC RESPONSES FROM BYSTANDERS

### 6.1 Empathetic Responses Evoked from Threat and Pain Perception

Two different hypotheses regarding the modulation of empathy on threat perception have been proposed in the literature. The "empathising hypothesis" claims that observation of another's pain promotes feelings of empathy, influencing prosocial behaviours towards the victim [85]. Meanwhile, the "threat value of pain hypothesis" states that the processing of painful and harmful stimuli may activate one's own threat-detection system, promoting defensive behaviours [86]. However, the characteristics of the threat and environmental contexts also play a crucial role in threat appraisal and eliciting defensive behaviours. For instance, seeing someone injured in a hospital, would be less alarming and may even incite prosocial behaviour. Meanwhile, witnessing an injured individual in the supermarket, may instead serve as a warning of impending danger, promoting threat detection over prosocial behaviours. Pain is commonly used within experimental contexts to evoke feelings of threat, due to their ability to induce defensive behaviours [86]. Thus, to gain a better understanding on the complex role of empathy within pain perception, we ought to investigate how empathy is measured and captured within the literature.

## 6.2 Brain Regions Implicated in Empathetic Responses to Observed Pain

Empathy is an understudied phenomena within threat perception research, despite the influence it has on the perception of another's pain. Pain networks are identified in threat perception, on the basis that pain, and threat are often linked, as individuals often deem potential pain as a threat and become threat averse. fMRI studies have identified two cortical and subcortical networks for pain empathy that are capable of quick bottom-up activations, which are required in quick attentional processing of threat [84]. The observation of an individual in a painful situation results in increased activation of the somatosensory cortex, evoking pain-related responses in the motivational-affective dimension of pain within oneself [87]. The prefrontal cortical circuitries have been identified to hold influence over high-level signals, that are based on predictions and prior experiences, substantiating the BIS and its mediation of defensive behaviours [84]. Meanwhile, the insula and cingulate cortices have also shown increased activation upon the observation of another's pain, within individuals with higher self-reported levels of empathy [88].

The amygdala is believed to be central in emotional empathy, due to considerable evidence of the amygdala playing a role in deciphering fear-related stimuli [89]. One perspective claims that the amygdala contributes to empathy by encoding the severity of threat [90]. This model states that fear-encoded stimuli reinforce the activation of the amygdala, with more fearful stimuli resulting in greater amygdala activity [89], as observed by the amygdala responding more actively to the increased intensity of fearful facial expressions [91]. In contrast, another model proposes that the amygdala does not encode the levels of distress, but rather redirects cognitive functions like attention towards the object's features, to disambiguate the stimuli [89]. As a result, increased amygdala activity is predicted for more ambiguous stimuli that requires a greater understanding of situational cues to decipher. Hence, this model predicts that once attention has been successfully redirected to relevant stimulus features, the amygdala is no longer utilised for emotion recognition [89].

In order to investigate the amygdala's role in the perception and identification of fear, when different contextual cues are present, a fMRI study was conducted utilising a region of interest (ROI) approach to identify amygdala responses to emotionally charged and emotionally ambiguous stimuli [89]. Results indicated that amygdala activity was associated with higher levels of fearful stimuli, reflecting the level of arousal incited by complex visual scenes. However, a lack of amygdala activity indicated it was not modulated by emotional ambiguity, during the discrimination between differing emotion context stimuli [89]. Hence, these results indicate that the amygdala plays an essential role in threat perception by encoding the severity of threat, from fearful expressions. These results are significant to models of empathy, as the recognition of fear in others is vital to empathetic behaviours, which are fundamental to prosocial behaviour [94].

Contextual cues also contribute to the degree of empathetic appraisal experienced from the perception of another's pain. Under differing social contexts, like cooperative versus competitive and friendly versus threatening, participants perceived another's pain differently [92]. These findings show that a combination of threatening and competitive stimuli resulted in increased arousal in participants, indicating the role of specific social circumstances in modulating neural responses to pain in others. This manipulation of social contexts creates a contrast in the probability and predictability of perceiving pain; hence, pain perception is more likely for the competitive/threatening context than the cooperative/friendly context [84]. As a result, contextual cues can modulate an observer's pain perception by mediating experienced arousal.

## 6.3 Methodological Limitations in the Study of Empathetic Appraisal from Pain Perception

Despite pain being a central component within threat, pain, and threat are not necessarily mutually exclusive, as the repercussions of threat need not solely be physical damage. The nature of threat can extend beyond the physical, manifesting as psychological threats. Hence, although pain is considered a threat, not all types of threat consist of pain. Consequently, when most studies within the literature use visual stimuli of painful scenarios, it only serves to inform the neuro-physical responses to pain, as opposed to the more generalised field of threat.

Furthermore, consolidating the dynamic nature of threat into static stimuli makes it hard to control what sensations participants might be experiencing. For example, a painful image might not necessarily incite feelings of empathy for another's pain, instead inciting feelings of disgust or shock, depending on the arousal intensity of visual stimuli.

Lastly, it is difficult to measure and analyse feelings of empathy, especially when the participants have no experimental opportunity to indicate this, through means of experimental outcomes such as prosocial behaviours or qualitative measurements such as personality questionnaires. Thus, to generalise findings to experienced threats, experiments should not only utilise ecologically valid and immersive stimuli, but also present participants with the option to respond empathetically, for statistical measure.

## 6.4 Empathetic Responses Evoked from Embodied Perspective Taking with VR

The ability to embody other perspectives with VR can capture the dynamic nature of threat to elicit empathetic responses, thus, influencing socio-behavioural processes. Interpersonal threat can be emulated by virtual agents, with emotional signals consisting of facial and bodily expressions [95]. VR simulations are used to evoke embodiment illusions, whereby one's sense of body ownership is manipulated by the perceptual illusion or the feeling of embodying in a non-human or virtual entity. Embodiment illusions have the potential to affect social cognition by influencing behavioural responses and perceptions [93].

Fearful expressions are more readily modulated within embodiment illusions than other emotional expressions [96]. This feat is observed by the perceptual illusion of the Visual Remapping of Touch effect, which is enhanced accuracy in tactile perception of one's face upon observation of another's face being touched [97]. However, this illusion solely facilitates fear expression recognition, unlike other expressions, implying that fear is perceived and processed differently [98]. Hence, perspective-taking of another's multisensory experiences, through embodiment, is speculated to facilitate regions of the somatosensory cortex that are related to fear perception [93]. However, unlike non-VR illusions, VR has been shown to be capable of promoting strong feelings of ownership, resulting in fronto-parietal activation [99], despite differing appearances of one's own body to the present avatar [100]. In comparison to narrative imagination tasks, VR has resulted in significantly higher degrees of the unity of senses and perspective taking [102]. Furthermore, VR has been dubbed as the "empathy machine" [101], showing profound effects of the alteration of one's self-bodily perception, affecting self-intrinsic socio-cognitive processes, which can be manipulated to solve issues within the real world [93].

The actions of violent offenders are postulated by theoretical models of aggression, whereby empathy deficits prevent offenders from envisioning their victims' perspectives [105]. Empathy is believed to be a skill that moderates the exercising of aggressive behaviours [103]. Hence, an inclination towards aggression could be due to deficits of cognitive empathy, the ability to understand another's emotions and mental state, and emotional empathy, the feeling of vicarious emotion states [104]. According to Blair's Violence Inhibition Mechanism model, one's empathetic response can be prevented by the poor recognition of another's expressed anger and fear [105]. This effect has been observed in male offenders who display a significantly lower ability to recognise fear in female faces, in comparison to non-violent controls, as measured by a Face-body Compound emotion recognition test [93]. This evidence suggests a differential pattern activation in emotional expression perception, which coincides with violent behaviour being associated with negative emotion recognition deficits and ambiguous emotional stimuli processing differences [106]. Consequently, VR is speculated to provide the opportunity for embodied perspective-taking, presenting a potential intervention for socio-affective impaired individuals.

The perspective-taking element of VR can influence the emotional recognition skills and physiological responses of individuals, to promote empathetic appraisal. Perspective-taking often goes together with high level empathetic processing, being fundamental to the understanding of another's perspective [105]. To induce a full-body illusion, a male domestic violence offender's body was substituted by a virtual female that was synchronised with their movements, while the VR paradigm subjected the participant to scolding and aggressive behaviours from a virtual male. This opportunity to adopt the victim's perspective through virtual embodiment in VR, resulted in offenders displaying an improved ability in recognising fearful female faces and were able to accurately classify them as fearful, rather than happy [93]. Furthermore, these offenders displayed stronger physiological responses like more pronounced heart rate deceleration (HRD), an anxiety marker of defensive and vigilant responses elicited from aversive stimuli [79], when observing explicit acts of aggression, such as when the virtual aggressor threw a phone on the floor, in comparison to implicit non-verbal cues of approaching the individual [107]. Increased HRD in response to threat, whilst embodying a virtual body has been positively correlated to body ownership feelings [107].

In a later fMRI study, an enhancement in activity within the Default Mode Network (DMN), which is involved in introspection and memory retrieval, was captured when offenders were subjected to violent aggression from the perspective of a victim [95]. In contrast, DMN activity was reduced when perceiving fearful victim expressions, indicating that the perception of fear as a discrete emotion was easier to recognise in comparison to emotionally ambiguous stimuli [96]. However, as this is a novel study introducing neuroimaging to quantify the effects of VR on emotional recognition, corresponding literature to substantiate this claim is limited.

Differing perspectives adopted within VR paradigms can influence the saliency of emotions in an observer, resulting in different affective, neural, and behavioural responses. When an individual embodies the direct victim of an assault, they experience stronger fear arousal [108]. Similarly, studies found that first-person perspective embodiment resulted in increased body ownership, fear arousal, physiological reactivity, amygdala activation and frontoparietal network synchronisation to predict another's actions, in comparison to third-person embodiment [109], [110]. However, when a third-person perspective is embodied, violence promotes stronger feelings of vicarious anger over fear [108].

### 6.5 Methodological Limitations from Studies Investigating Embodied Perspective Taking in VR

Despite embodied perspective taking enabling the study of affective and cognitive mechanisms, most of the aforementioned work has been centred in extreme populations, with work focusing on the rehabilitation of offenders. This may limit the generalisability of results, due to potential confounding factors like substance abuse or abnormal mental processing. Regardless of the use of an extreme population, findings from these studies provide insight on the cognitive mechanisms implicated in embodied perspective taking, by empathy and emotion recognition Furthermore, variation in intensity of conflict, the study of these uniform cognitive mechanisms can be applied to the general population, due to the nature of conflict being socially pervasive.

Regardless of these limitations, the studies provide foundational evidence for VR being used to facilitate the modulation of neuropsychological deficits to minimise the repercussions of poor emotion recognition. Additionally, the extensive nature of the VR environment can be utilised to provide real-time feedback to strategically shape and modulate performance of the participant to accomplish specific goals [111]. The use of VR to train teachers showcases the benefits from real-time feedback, whereby, the algorithms and situational elements can adjust to the behaviours of the teachers, to enhance the skill tested while promoting new skills [107]. Feedback and rewards within a virtual environment have the potential to increase engagement within the environment and guide participants along to the objectives of their tasks [108]. The measurement and modulation of behavioural responses occurs with ease in VR contexts, whilst maintaining ecologically relevant settings. All these characteristics of VR promote the level of experimental control necessary for the measurement, analysis, and replication of threat perception behavioural studies.

## 6.6 Threat Perception Evoking Empathetic Attitudes, Resulting in Prosocial Behaviours

Empathy is the key mechanism that drives selfless decisions, due to the broader evolutionary framework influencing situation-specific behaviours, which can alleviate the distress of others [40]. This inclination to resonate with the emotions of another, is believed to have been adapted from the mother-offspring bond, enabling the quick recognition of vulnerability and distress within another [40]. Multiple studies have demonstrated that individuals of greater empathetic concern are more likely to display prosocial behaviours such as by giving up monetary reward and taking pain-inducing electric shocks to reduce another's suffering [40]. Hence, empathy is crucial to the exercising of prosocial behaviours, which ultimately minimises harm and reduces the likelihood of the bystander effect.

Threat imminence can modulate empathetic appraisals of prosocial behaviours. A study conducted required participants to make trial-by-trial decisions as to whether they would aid a co-participant in avoiding an aversive shock, at the potential expense of receiving a shock [40]. The independent variable was the threat's imminence; thus, decisions were prompted when threat was either imminent or not. The imminence of threats resulted in faster prosocial responses and increased heart rates [40]. Therefore, these findings suggest that imminent threats promote defensive states present in fight or flight behaviours to facilitate prosocial decisions, towards other individuals experiencing distress. As a result, these results imply that the fight or flight nature of threat perception is linked to one's choice to engage in a prosocial nature.

Some studies have taken an appraisal approach to prosocial motivations, whereby affective states influence different methods of cognition [112]. Within this framework, compassion is distinctive from empathy, as it is more quantitatively measurable and is defined by prosocial care motivation and empathic recognition of another's suffering [113]. Hence, a subfield of literature distinguishes empathy from compassion, due to compassion encompassing the prosocial action taken to reduce the suffering of another [114]. This distinction prioritises the study of compassion over empathy, as it is believed that compassion drives prosocial behaviours [30].

To investigate whether acts of compassion operate similarly to empathy by activating empathy networks, neurobiological measures must be taken. Due to compassion being defined as the ability to not only empathetically recognise another's suffering but also the displaying of prosocial behaviours, it implies that the activated neural networks should be like those activated by empathy. A fMRI study conducted predicted that acts of compassion towards a vulnerable individual, would activate regions of the empathy network like the anterior insula, and inferior frontal gyrus (IFG) and midbrain PAG [115]. Self-report data was collected to measure sentiments of compassion and distress after the presentation of each stimulus. Despite sharing activation of the PAG and IFG, key regions prevalent in the empathy network like the insula, medial prefrontal cortex (mPFC) or human temporal polar cortex were not activated when participants were compassionate [115].

The difference in neural network activation could be attributed to the emotion induction method utilised. Participants were required to continuously self-monitor, self-evaluate and appraise their emotions [116]. As a result, the self-reflective nature of the task might have stimulated evaluative regions like the insula and mPFC, hence affecting the qualitative differences between stimuli conditions [115]. These findings contribute towards the understanding of neural pattern activation similarities from acts of compassion and empathy. However, a further investigation on empathetic appraisal resulting in prosocial actions may require more ecologically valid methods of measurement to preserve the integrity of findings.

Empathetic behaviours can reduce the probability of threat confrontation, by providing a greater insight into another's perspective to aversive stimulus [31]. To promote threat confrontation reduction, empathetic appraisal needs to complement the reinforcement of memories learnt through first-hand aversive experiences [117]. A study investigated whether the observation of another's threat response would reinstate one's own, prior learnt defensive responses to a similar threat [117]. The experiment was split into two conditions, whereby observers within both conditions watched a visual recording of a demonstrator react to three, unannounced electrical shocks. Following this, all participants were presented with similar scenarios as the individuals presented in the videos and skin conductance was measured in both phases of the experiment. However, unlike the demonstrator videos, participants did not receive a shock, yet they displayed elevated physiological responses to threat-conditioned cues, after the observation of another's aversive response towards threat stimuli. Participants also displayed increased arousal in contexts perceived as dangerous [117]. Thus, indicating that the observation of another's threat responses has the potential to reinstate threat responses in an observer [117].

As a result, being empathetic to another's experience with aversive stimuli, can serve to inform one's own affective experience with the stimuli. Consequently, this study substantiated that vicarious threat learning can recover associations shaped by the observation of another's aversive experiences [117]. Furthermore, the effect of empathetic appraisal promotes vicarious learning through the observation of another and one's own first-hand experiences [117].

# 7 FINAL REMARKS

## 7.1 Conclusions, Challenges and Future Research Directions

This paper has systematically reviewed the existing threat perception literature noting the influence of threat on emotion and motor systems. The analysis of number relevant articles brought rise to the following conclusions:

*"Are fear and aggression linked within threat? Are fight or flight behaviours the result of this relationship?"*

Fear and aggression are linked through the fight of flight phenomena, as observed by specific neural networks, allowing these emotion perception cues to influence corresponding defensive action systems. For instance, the amygdala was attributed to the processing of affective stimuli, with more amygdala activity was associated with higher levels of both fearful and aggressive stimuli, reflecting the level of arousal felt [89]. Meanwhile, a rage circuit comprising of the PAG was theorised and its activity is present in the face of threat [64].

Fight or flight behaviours appear to be evoked from sensations of fear and aggression. Purposeful and dynamic displays of aggression were correlated with high valence arousal ratings from physiological measurements [64]. Furthermore, fear is commonly correlated with aversive behaviours, as measured by startle response, within the literature [22]. Similarly, aggressive behaviours are often associated with approach motivation, comprising of confrontational behaviours [22]. Physiological sensations are crucial to the promotion of feelings of aggression and fear, due to subcortical regions of the brain promoting reflexive defensive responses to threat [25]. Hence, indicating a relationship between affective, physiological states and fight or flight behaviours.

*" Can biological human motion portray and capture measurable affective states from the fight or flight phenomenon?"*

The literature discussing the role of motor responses in capturing threat perception are two-fold, firstly, there has been studies utilising motion detection to infer cognitive processes of emotion or threat perception. This ranges from the use proxemics in VR to measure empathetic behaviours [69] and the use of classification algorithms to detect emotion states from body movements [61]. Sensations of fear from an imminent threat have also been correlated with one's own freezing behaviours [78]. Meanwhile, the perception of motor responses can also mediate empathetic appraisal towards virtual agents [74] and perceptions of threat from the induction of feelings of fear and aggression [64].

Cognitive flexibility mediates the adaptability of motor behaviours, in the face of threat. Cognitive flexibility promotes an attention bias towards threat within one's environment. This process can mediate the relationship between fear and aggression, resulting in corresponding fight or flight motor behaviours. This attention bias can also be the result of body immobility, evoked from fearful stimuli [28]. As a result, the perception of threat is prioritised over neutral cues, as observed by neural networks specialised in the perception of fearful body cues and expressions [28]. The perception of another's fearful body cues is prioritised in perception as visual stimuli. The distinction between fear and neutral stimuli is observed by the increased sensitivity of certain brain regions, such as the EBA to fear stimuli.

The measurement of motor responses in realistic scenarios brings about challenges around replication, validity, and ethics. Thus, VR was proposed as a method to extract emotions from motor movements with proxemic measures. With a greater degree of experimental control, VR can give real-time feedback to participants, while ensuring consistency within social interactions with an artificial confederate. The combination of VR and motion tracking presents an opportunity to study social interaction, alongside kinematics differences within varying emotional scenarios.

There is a lack of literature found that identified the effect of aggression on motor behaviours, due to most research found within our search, focusing on fear elicitation. Hence, future research ought to investigate whether there is a link between aggression as an affective state and the potential for corresponding fight or flight defensive behaviours.

*"What factors of threat perception modulates empathetic responses and how are these measured by affective computing?"*

Within our systematic search the main two methods of using VR to evoke empathetic responses in observers, was from the perception of another's pain and embodied perspective taking. Perceiving another's pain can be a strong motivator for empathetic behaviour, with immediacy contextual cues heavily influencing empathetic responses. However, pain is not mutually exclusive to threat, despite fear being a commonly shared affective state between them. Thus, future research investigating the effect of threat perception on empathetic responses ought to investigate impending threat scenarios that are not solely reliant on physical pain.

Meanwhile, embodied perspective-taking results in higher displays of emotion congruency and the recognition of fearful expressions [93]. The limited literature supports these findings, depicting empathetic responses as a result in significant neurological activity changes [95]. Despite this field in the literature being relatively novel, it shows promising results for the study of varied perspective taking on empathetic appraisal.

*"How do empathetic appraisals result in prosocial behaviour, and has this been measured by affective computing?"*

Empathetic appraisals have the potential to evoke prosocial behaviour, as fight or flight defensive states can facilitate prosocial decisions towards individuals in need (seen by pain perception and embodied perspective-taking). The imminence of threat has been found to modulate empathetic appraisals, towards another's suffering, promoting behaviours to alleviate another's distress [40].

A prominent challenge within the literature was the variability in definitions used for empathetic states. These differing definitions draw attention to varying methods of assessment. For instance, the measurement of affective empathy, as observed by empathic mimicry, would prioritise measurements of the vicarious experience of another's emotions. Meanwhile, cognitive empathy, which consists of higher-level cognitive processing, refers to mentalising how others feel and could be measured by more qualitative methods, such as trait empathy self-assessment scales. Consequently, the variation in findings exploring empathetic networks, can be explained by non-overlapping parts of empathy, as posited by differing definitions, being captured [118]. Furthermore, there has been a new school of thought, attempting to distinguish empathy from compassion, however these differences may be due to methodological limitations, rather than definitions [30]. Hence, due to the diverse definitions and corresponding measurements of empathy, diverging opinions and subdomains regarding empathetic behaviour create a complex literature.

Beyond consolidating diverse definitions, future affective research ought to provide more clarity for affective states, by conducting more sensitive continuous measures of emotional reactivity, as opposed to the current literature taking intermittent measurements. Several suggestions have been made within the literature to improve the quality of affective research. Firstly, due to greater skin conductance being identified from the viewing of emotional images, it can be used to identify the period of peak emotional arousal, identifying emotion induction and its impact on defensive behaviours [68]. Secondly, the emotional processing theory can be referenced when designing mood induction procedures, to aid the disambiguation between acute and sustained reactivity [25]. It is an organisational framework aiming to increase adaptive responses across cognitive, behavioural, and physiological domains [119]. Therefore, referencing this model may strengthen the validity, potency, and generalisability of future results, by focusing on the stimulus characteristics of the scenario, over types of responses [25-]. Henceforth, when it comes to measuring affective responses within situations that require split-second responses, it is essential for precisive methods to be utilised, to not only capture, but maintain the validity of dynamic affective responses.

Within this systematic review, limitations consisting of the varied definitions of affective states, and the corresponding different methods of measurement were identified, contributing to a complex literature to streamline. Increasing varied methods of measurement from a lack of definition consensus, can explain the lack of overlapping findings and ought to be considered by future studies. Lastly, this review proposed the use of VR, as observed by the increased use in more recent studies, as a more precise and ecologically method of measuring behavioural responses to threat perception.

In conclusion, this systematic review assessed the existing literature to identify the relationship between fear and aggression, and fight or flight defensive behaviours, when in the presence of threat. Due to the nature of fight or flight behaviours, this review aimed to investigate role of biological motion in relation to fear or aggression, as implicated by the fight or flight phenomenon. This review paper also summarised the factors that influence threat perception on empathetic appraisals, and the use of affective computing to measure these responses. Lastly, empathetic appraisals were explored in relation to inducing prosocial behaviour, and whether the existing methods used to capture these behaviours are sufficient.

The strengths and limitations of methods and definitions identified within the literature were elaborated on. To our knowledge, this is the first systematic review of its kind that comprehensively explores the multifaceted nature of threat perception, from affective states to defensive motor behaviours, with the goal of harnessing this relationship for the societal benefits of modulating prosocial behaviours from empathetic appraisal.

## ACKNOWLEDGMENT

This work was supported by the UKRI centre for Doctoral Training in Socially Intelligent Artificial Agents, Grant Number EPS02266X1. This work was also supported by the Engineering and Physical Sciences Research Council, Grant Number EPW01212X1 and Royal Society, Grant Number RGSR2212199.

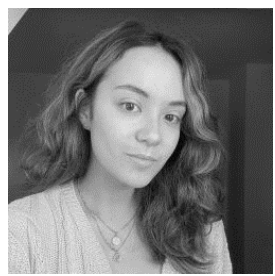

**Elizabeth M. Jacobs** received a bachelor's in science, in the field of Psychology, from the University of York, in 2021. She is a current PhD student in the integrated study of psychology and computing science, at the University of Glasgow. She is a research student within the SOCIAL AI CDT, which is funded by the UKRI. Her research interests include emotion science, non-verbal behavioural analysis and artificial intelligence.

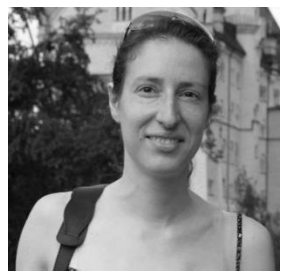

**Dr Fani Deligianni** holds a PhD in Medical Image Computing (Imperial College London), an MSc in Advanced Computing (Imperial College London), an MSc in Neuroscience (University College London) and a MEng (equivalent) in Electrical and Computer Engineering (Aristotle University, Greece). She developed sophisticate computational approaches in machine learning, statistics and network analysis for the investigation of human brain structure and function. Recently, her work is focused on human motion analysis with wearable sensors and single rgb(d) camera for healthcare applications.

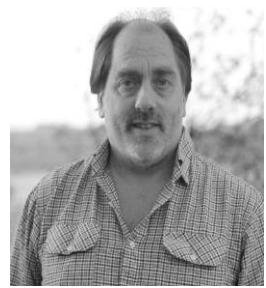

**Professor Frank Pollick** obtained BS degrees in physics and biology from MIT in 1982, an MSc in Biomedical Engineering from Case Western Reserve University in 1984 and a PhD in Cognitive Sciences from The University of California, Irvine in 1991. Following this he was an invited researcher at the ATR Human Information Processing Research Labs in Kyoto, Japan from 1991-97. He is interested in the perception of human movement and the cognitive neural processes that underlie our abilities to understand the actions of others.